\documentclass[preprint,12pt,sort&compress]{elsarticle}
\usepackage{amstext}
\usepackage{amsmath}
\usepackage{amssymb}
\usepackage{graphicx}
\usepackage{latexsym}
\usepackage{amssymb}

\journal{Annals of Physics}

\begin{document}

\begin{frontmatter}



\title{Distribution of Standard deviation of an observable among superposed states}


\author{Chang-shui Yu$^*$}
\cortext[cor1]{Corresponding author. Tel: +86 41184706201}
\ead{ycs@dlut.edu.cn}
\author{Ting-ting Shao, Dong-mo Li}

\address{School of Physics and Optoelectronic Technology, Dalian University of
Technology, Dalian 116024, China}

\begin{abstract}
The standard deviation (SD) quantifies the spread of the observed values on
a measurement of an observable. In this paper, we study the distribution of
SD among the different components of a superposition state. It is found that
the SD of an observable on a superposition state can be well bounded by the SDs
of the superposed states. We also show that the  bounds
also serve as good bounds on  coherence of a superposition state.  As a further generalization, we give an alternative definition of incompatibility of two observables subject to a given state and show how the incompatibility subject to a superposition state is distributed.\end{abstract}

\begin{keyword}
Standard deviation \sep quantum coherence \sep incompatibility \sep quantum superposition
\PACS{03.67.Mn\sep 03.65.Ta \sep03.65.Ud}


\end{keyword}

\end{frontmatter}


\section{ Introduction}

Quantum superposition is the most fundamental feature of quantum mechanics.
Almost all the intriguing quantum phenomena are directly or indirectly
related to quantum superposition. For example, it is the necessary factor
for the interference of microscopic particles. In particular, combined with
the tensor product structure of quantum state space, it can produce the most
remarkable quantum phenomenon-----quantum entanglement which forms an
important physical resource in quantum information processing \cite%
{Horodecki}. However, the superposition in quantum mechanics does not always
play the expected role. It could also lead to the coherent destruction. The
obvious example is the vanishing entanglement for the superposition of two
Bell states with equal amplitudes. So a natural question is how the
entanglement is distributed among the different components of the
superposition state or whether we can give a reference evaluation of the
entanglement for the superposition state based on the entanglement of every
component? This question was first addressed in Ref. \cite{LPS}, by Linden,
Popescu and Smolin who found that the entanglement of superposition states
(measured by the von Neumann entropy of reduced density matrix) was upper
bounded in terms of the entanglement of each component. From then on, the
entanglement of superposition states have attracted wide interests ranging
from different bounds to various entanglement measures \cite{Yu, Gour, Cav,
Cerf, Fan, Xiang,Yu2,Yu3,akh,par,bhar,bhar2,Ma}. In addition, entanglement
has been extensively studied in various systems \cite{r7} such as graphenes 
\cite{r6} and optomechanical systems \cite{r5} , and even in living object 
\cite{r3,r4}. The experimental preparation of quantum entanglement is also
progressing fast \cite{r8,r9,r10,r11}. 
\begin{figure}[tbp]
\centering
\includegraphics[width=0.8\columnwidth,height=2.35in]{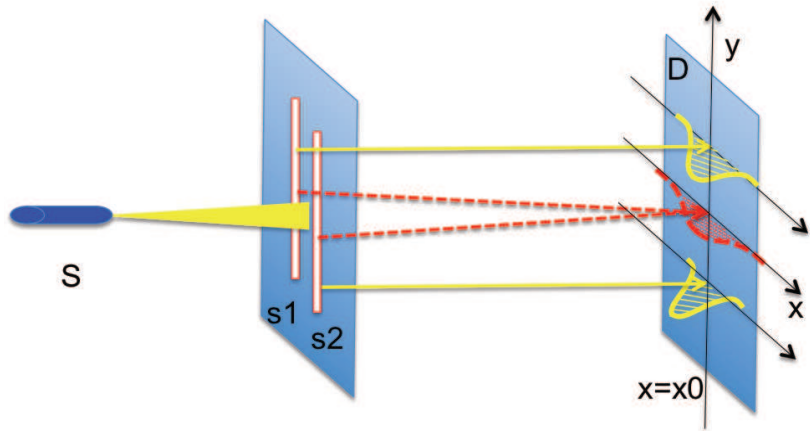} \centering
\caption{(Color online.)Diagrammatic sketch of double-slit experiment. The
photon emitted from the coherent light source $S$ passes through two (or one
of) slits $s1$ and $s2$ and is detected at position $x_0$ on the screen $D$.
Along $y$ axis the SDs are shown at $x=x_0$ with different slits open. The
yellow lines corresponds to the SD with only one slit open and the red line
represents the SD with both slits open.}
\end{figure}

In practical physics, the measurement is the absolute requisite which allows
us to know the objective world \cite{Busch0}. But any measurement is
imperfect, so one has to perform repeated measurements to reduce the
distance between the average result and the real value, that is, the
measurement error. The standard deviation (SD) which quantifies the spread
of the observed values on a measurement of an observable, is usually used to
characterize the measurement error \cite{Nl}. However, in the quantum world,
besides the classical measurement errors, the quantum nature results in that
measurements of an observable on the same quantum state don't generally
produce the same measurement value. So the SD also characterizes the
essential uncertainty of a single measurement of an observable subject to a
certain state. It further plays the important role in the remarkable
Heisenberg's uncertainty principle (see Ref. \cite{Busch} and the references
therein). It is also worth pointing out that the SD of observable in an
ensemble has been used to distinguish ensembles with the same density matrix 
\cite{r12}. A related study can be also found in \cite{r13}. Since the
superposition of states is a universal phenomenon in the quantum world, 
\textit{how is the SD distributed among the different components of the
superposition states? Or how can we effectively evaluate the SD of the
superposition state in terms of the SD of every component?} For example, in
a double-slit experiment shown in Fig. 1, suppose only one slit is open
once, we can detect photon at a given position on the screen.
Correspondingly, one can obtain the SDs of the position operator subject to
each slit, respectively. Can we evaluate the SD at the same position in
terms of the SD corresponding to each slit when both slits are open? In
addition, since the usual uncertainly relation (or the incompatibility of
observables) is given by the SDs of two incompatible observables, answering
such a question could also provide a significant understanding on how the
superposition influences the uncertainty relation.

In this paper, we answer the above questions by a general treatment in
mathematics. It is found that the SD of the superposition state is well
upper and lower bounded by the SDs of the superposed components. In
addition, based on the formally consistent definitions of the SD and the
coherence, it is interesting that the presented bounds on the SD can also
serve as good bounds on coherence of the superposition state. That is, the
bounds of the SD provide formally unified bounds on the SD and the
coherence. Considering the connection with the uncertainty relation, we
present an alternative incompatibility measure of two observables subject to
a given state. It is shown that the bounds on the SD can also induce good
bounds on the incompatibility of superposition states. The paper is
organized as follows. In Sec. II, we present our main result about the
bounds on the SD. In Sec. III, we provide the unified bounds on the
coherence. In Sec. IV, we present the incompatibility measure of two
observables and give its bounds. The conclusion is drawn finally.

\section{Bounds on the standard deviation}

To begin with, let's consider an observable $A$ which is measured on a state 
$\left\vert \psi \right\rangle $. The SD is defined by%
\begin{equation}
\Delta _{\psi }A=\sqrt{\left\langle A^{2}\right\rangle -\left\langle
A\right\rangle ^{2}},  \label{def1}
\end{equation}%
where the subscript $\psi $ denotes that the expectation value $\left\langle
X\right\rangle =\left\langle \psi \right\vert X\left\vert \psi \right\rangle 
$ for any observable $X$ is taken on the state $\left\vert \psi
\right\rangle $. In the following, we also use $\left\langle
X\right\rangle_j $ to denote $\left\langle\psi_j \right\vert
X\left\vert\psi_j\right\rangle$ for some labeled state $\left\vert\psi_j%
\right\rangle$. Thus we suppose $\left\vert \psi \right\rangle $ is a
superposition state, we aim to find how $\Delta _{\psi }A$ is bounded by the
SDs of the superposed components. In this sense, we look for the bounds that
can be expressed by the quantities directly related to the SD of every
component instead of potentially tighter bounds given by other irrelevant
quantities.

\noindent \textbf{Theorem 1.}-\textit{Let the observable }$A$ \textit{be
measured on the superposition state} $\left\vert \psi \right\rangle
=\sum\limits_{i=1}^{N}\alpha _{i}\left\vert \psi _{i}\right\rangle $ \textit{%
with} $\sum\limits_{i=1}^{N}\left\vert \alpha _{i}\right\vert ^{2}=1$\textit{%
\ and }$\left\Vert \left\vert \psi _{i}\right\rangle \right\Vert =1$\textit{%
, the SD }$\Delta _{\psi }A$\textit{\ can be bounded as}%
\begin{equation}
\mathcal{B}_{L}\leq \left\Vert \left\vert \psi \right\rangle \right\Vert
^{2}\Delta _{\tilde{\psi}}^{2}A\leq \mathcal{B}_{U}  \label{bds}
\end{equation}%
\textit{where }$\left\vert \tilde{\psi}\right\rangle =\frac{1}{\left\Vert
\left\vert \psi \right\rangle \right\Vert }$ $\left\vert \psi \right\rangle $%
\textit{\ with }$\left\Vert \left\vert \psi \right\rangle \right\Vert $%
\textit{\ denoting the l}$_{2}$\textit{\ norm of a vector,} 
\begin{equation}
\mathcal{B}_{L}=\max \{0,b_{L}\}  \label{lb}
\end{equation}%
\textit{with}%
\begin{equation}
b_{L}=\sum\limits_{i=1}^{N}\left\vert \alpha _{i}\right\vert ^{2}\Delta
_{_{\psi _{i}}}^{2}\left( A\right) -E_{+}(A)-F(A),  \label{lbc}
\end{equation}%
\textit{and} 
\begin{equation}
\mathcal{B}_{U}=\sum\limits_{i=1}^{N}\left\vert \alpha _{i}\right\vert
^{2}\Delta _{_{\psi _{i}}}^{2}\left( A\right) -E_{-}(A)+F(A).  \label{ub}
\end{equation}%
\textit{with } 
\begin{equation}
E_{\pm }(A)=\frac{\left( \left\vert \sum\limits_{i=1}^{N}\left\vert \alpha
_{i}\right\vert ^{2}\left\langle A\right\rangle _{i}\right\vert \pm
\sum\limits_{i,j=1,i<j}^{N}2\left\vert \alpha _{i}\alpha _{j}\right\vert 
\sqrt{\left\vert \left\langle A\right\rangle _{i}\left\langle A\right\rangle
_{j}\right\vert }\right) ^{2}}{\left\Vert \left\vert \psi \right\rangle
\right\Vert ^{2}}+\sum\limits_{i=1}^{N}\left\vert \alpha _{i}\right\vert
^{2}\left\langle A\right\rangle _{i},  \label{epm}
\end{equation}%
\textit{and} 
\begin{equation}
F(A)=\sum\limits_{i,j=1,i<j}^{N}2\left\vert \alpha _{i}\alpha
_{j}\right\vert \sqrt{\left( \Delta _{_{\psi _{i}}}^{2}\left( A\right)
+\left\langle A\right\rangle _{i}^{2}\right) \left( \Delta _{_{\psi
_{j}}}^{2}\left( A\right) +\left\langle A\right\rangle _{j}^{2}\right) }.
\label{fa}
\end{equation}

\noindent \textbf{Proof.} Based on the definition of the SD of $A$ on $%
\left\vert \psi \right\rangle $, we can have 
\begin{eqnarray}
\Delta _{\tilde{\psi}}^{2}\left( A\right)  &=&\left\langle
A^{2}\right\rangle _{\tilde{\psi}}-\left\langle A\right\rangle _{\tilde{\psi}%
}^{2}  \notag \\
&=&\frac{1}{\left\Vert \left\vert \psi \right\rangle \right\Vert ^{2}}%
\left\langle A^{2}\right\rangle _{\psi }-\frac{1}{\left\Vert \left\vert \psi
\right\rangle \right\Vert ^{4}}\left\langle A\right\rangle _{\psi }^{2},
\label{p1}
\end{eqnarray}%
where the subscript denotes the expectation value on the corresponding
state. Expanding $\left\langle A^{2}\right\rangle _{\psi }$, one arrives at%
\begin{equation}
\left\langle A^{2}\right\rangle _{\psi }=\sum\limits_{i=1}^{N}\left\vert
\alpha _{i}\right\vert ^{2}\left\langle A^{2}\right\rangle
_{i}+\sum\limits_{i,j=1,i\neq j}^{N}\alpha _{i}^{\ast }\alpha
_{j}\left\langle A^{2}\right\rangle _{ij},  \label{p2}
\end{equation}%
with $\left\langle X\right\rangle _{ij}=\left\langle \psi _{i}\right\vert
X\left\vert \psi _{j}\right\rangle $ and $\left\langle X\right\rangle
_{j}=\left\langle \psi _{j}\right\vert X\left\vert \psi _{j}\right\rangle $
for any observable $X$. Substitute Eq. (\ref{p2}) into Eq. (\ref{p1}), we
will have 
\begin{eqnarray}
&&\left\Vert \left\vert \psi \right\rangle \right\Vert ^{2}\Delta _{\tilde{%
\psi}}^{2}\left( A\right)   \notag \\
&=&\sum\limits_{i=1}^{N}\left\vert \alpha _{i}\right\vert ^{2}\Delta
_{_{\psi _{i}}}^{2}\left( A\right) +\sum\limits_{i=1}^{N}\left\vert \alpha
_{i}\right\vert ^{2}\left\langle A\right\rangle
_{i}+\sum\limits_{i,j=1,i\neq j}^{N}\alpha _{i}^{\ast }\alpha
_{j}\left\langle A^{2}\right\rangle _{ij}  \notag \\
&-&\frac{1}{\left\Vert \left\vert \psi \right\rangle \right\Vert ^{2}}\left(
\sum\limits_{i=1}^{N}\left\vert \alpha _{i}\right\vert ^{2}\left\langle
A\right\rangle _{i}+\sum\limits_{i,j=1,i\neq j}^{N}\alpha _{i}^{\ast }\alpha
_{j}\left\langle A\right\rangle _{ij}\right) ^{2}.  \label{p3}
\end{eqnarray}%
For an observable $X$, one can always find%
\begin{equation}
\alpha _{i}^{\ast }\alpha _{j}\left\langle X\right\rangle _{ij}+\alpha
_{i}\alpha _{j}^{\ast }\left\langle X\right\rangle _{ji}\leq 2\left\vert
\alpha _{i}\alpha _{j}\right\vert \sqrt{\left\vert \left\langle
X\right\rangle _{i}\left\langle X\right\rangle _{j}\right\vert },  \label{p4}
\end{equation}%
which is based on the triangular inequality $\left\vert \sum
a_{i}\right\vert \leq \sum \left\vert a_{i}\right\vert $ for numbers $a_{i}$
and Cauchy-Schwarz inequality. Eq. (\ref{p3}) will become%
\begin{eqnarray}
&&\left\Vert \left\vert \psi \right\rangle \right\Vert ^{2}\Delta _{\tilde{%
\psi}}^{2}\left( A\right)   \notag \\
&\geqslant &\sum\limits_{i=1}^{N}\left\vert \alpha _{i}\right\vert
^{2}\Delta _{_{\psi _{i}}}^{2}\left( A\right)
+\sum\limits_{i=1}^{N}\left\vert \alpha _{i}\right\vert ^{2}\left\langle
A\right\rangle _{i}-\sum\limits_{i,j=1,i<j}^{N}2\left\vert \alpha _{i}\alpha
_{j}\right\vert \sqrt{\left\langle A^{2}\right\rangle _{i}\left\langle
A^{2}\right\rangle _{j}}  \notag \\
&-&\frac{\left( \left\vert \sum\limits_{i=1}^{N}\left\vert \alpha
_{i}\right\vert ^{2}\left\langle A\right\rangle _{i}\right\vert
+\sum\limits_{i,j=1,i<j}^{N}2\left\vert \alpha _{i}\alpha _{j}\right\vert 
\sqrt{\left\vert \left\langle A\right\rangle _{i}\left\langle A\right\rangle
_{j}\right\vert }\right) ^{2}}{\left\Vert \left\vert \psi \right\rangle
\right\Vert ^{2}},  \label{p5}
\end{eqnarray}%
where we use the positivity property of $A^{2}$. Since Eq. (\ref{p5}) could
be negative, but $\left\Vert \left\vert \psi \right\rangle \right\Vert
^{2}\Delta _{\tilde{\psi}}^{2}\left( A\right) $ is never negative, we have
to take the maximum between zero and the lower bound given by Eq. (\ref{p5}%
). Similarly, one can also obtain the upper bound from Eq. (\ref{p3}) as 
\begin{eqnarray}
&&\left\Vert \left\vert \psi \right\rangle \right\Vert ^{2}\Delta _{\tilde{%
\psi}}^{2}\left( A\right)   \notag \\
&\leq &\sum\limits_{i=1}^{N}\left\vert \alpha _{i}\right\vert ^{2}\Delta
_{_{\psi _{i}}}^{2}\left( A\right) +\sum\limits_{i=1}^{N}\left\vert \alpha
_{i}\right\vert ^{2}\left\langle A\right\rangle
_{i}+\sum\limits_{i,j=1,i<j}^{N}2\left\vert \alpha _{i}\alpha
_{j}\right\vert \sqrt{\left\langle A^{2}\right\rangle _{i}\left\langle
A^{2}\right\rangle _{j}}  \notag \\
&-&\frac{\left( \left\vert \sum\limits_{i=1}^{N}\left\vert \alpha
_{i}\right\vert ^{2}\left\langle A\right\rangle _{i}\right\vert
-\sum\limits_{i,j=1,i<j}^{N}2\left\vert \alpha _{i}\alpha _{j}\right\vert 
\sqrt{\left\vert \left\langle A\right\rangle _{i}\left\langle A\right\rangle
_{j}\right\vert }\right) ^{2}}{\left\Vert \left\vert \psi \right\rangle
\right\Vert ^{2}}.  \label{p6}
\end{eqnarray}%
Here we use the triangular inequality $\left\vert a_{1}+a_{2}\right\vert
\geqslant \left\vert \left\vert a_{1}\right\vert -\left\vert
a_{2}\right\vert \right\vert $ for two numbers $a_{1}$ and $a_{2}$.
Substitute $\left\langle A^{2}\right\rangle _{i}=\Delta _{_{\psi
_{i}}}^{2}\left( A\right) +\left\langle A\right\rangle _{i}^{2}$ into Eq. (%
\ref{p5}) and Eq. (\ref{p6}), one will immediately obtain the bounds given
in Eq. (\ref{lb}) and Eq. (\ref{ub}), which completes the proof.$\hfill $ $%
\blacksquare $

From Theorem 1, one can find that whether the equality in Eq. (\ref{bds}) is
achieved strongly depends on the considered observable $A$ and the
superposed states. It can be shown that the equality saturates for two
superposed states once one of the states happens to be the eigenvector of $A$
corresponding to its zero eigenvalue. In order to further show how tight the
bounds are, we randomly generate a $4\times 4$ Hermitian operator 
\begin{equation}
A=\left( 
\begin{array}{cccc}
-1.3343 & -0.7485 & -0.5932 & 0.1623 \\ 
-0.7485 & 0.2060 & -0.0115 & 0.9184 \\ 
-0.5932 & -0.0115 & -0.3338 & 0.3307 \\ 
0.1623 & 0.9184 & 0.3307 & 1.2613%
\end{array}%
\right)  \label{mat}
\end{equation}%
and two 4-dimensional quantum states $\left\vert \psi _{1}\right\rangle =%
\left[ 
\begin{array}{cccc}
0.5506 & 0.3628 & 0.6016 & 0.4509%
\end{array}%
\right] ^{T}$ and $\left\vert \psi _{2}\right\rangle =\left[ 
\begin{array}{cccc}
0.3511 & 0.4912 & 0.5296 & 0.5958%
\end{array}%
\right] ^{T}$. The superposition state is given by $\left\vert \psi _{\pm
}\right\rangle =x\left\vert \psi _{1}\right\rangle \pm \sqrt{1-x^{2}}%
\left\vert \psi _{2}\right\rangle $. We plot the upper and lower bounds and
the SDs of the states $\left\vert \psi _{\pm }\right\rangle $ in Fig. 2 and
Fig. 3. One can find from Fig. 2 that even though the lower bound is not so
tight as the upper bound for $\left\vert \psi _{+}\right\rangle $, we still
think the lower bound is also a tight bound, because in Fig. 3, with the
same expressions of the bounds, the lower bound is much tighter than the
upper bound for $\left\vert \psi _{-}\right\rangle $.

\begin{figure}[tbp]
\centering
\includegraphics[width=0.8\columnwidth,height=2.35in]{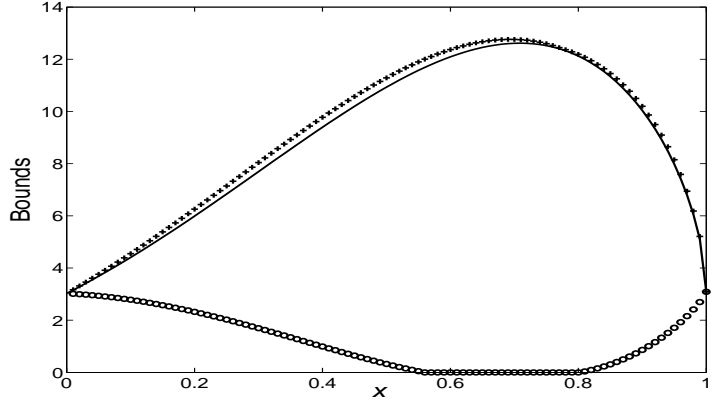} \centering
\caption{The upper and lower bounds and the SD of $\left\vert\protect\psi%
_+\right\rangle$ versus $x$. The '+' line corresponds to the upper bound,
the 'o' line corresponds to the lower bound and the solid line is the SD of $%
\left\vert\protect\psi_+\right\rangle$.}
\end{figure}
\begin{figure}[tbp]
\centering
\includegraphics[width=0.8\columnwidth,height=2.35in]{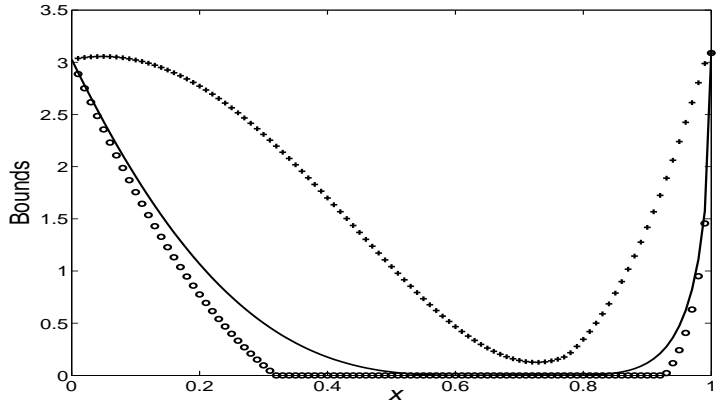} \centering
\caption{The upper and lower bounds and the SD of $\left\vert\protect\psi%
_-\right\rangle$ versus $x$. The line styles are defined similar to Fig. 2.}
\end{figure}

\section{Bounds on the coherence}

Quantum coherence stemming from quantum superposition of states is the most
fundamental feature of quantum mechanics. Recently, quantitative theory that
captures the resource character has been developed \cite{coh1}, even though
quantum coherence has been widely applied \cite{coh2,coh3,coh4}. It pointed
out that the good coherence measure should satisfy the following three
conditions \cite{coh1}: 1) Vanishing for incoherent states; 2) Not
increasing under incoherent operations; 3) Not increasing under mixing of
states. In fact, quantum coherence is also the essence of interference
phenomena, which shows that no interference could be revealed by the
observable if the observable commutes with the density matrix \cite{coh2}.
Here we would like to say that the quantum interference has also been
extensively studied in \cite{r1}, and resulted in the new concept of duality
quantum computers, which has found striking advantage in the scaling of
precision in quantum simulation \cite{r2}. Based on the commutation
property, an interesting coherence measure, \textit{K}-coherence, employing
the skew information has been raised \cite{cohs}. For a state $\varrho $,
the \textit{K}-coherence subject to an observable $K$ is defined by 
\begin{equation}
I\left( \varrho ,K\right) =-\frac{1}{2}Tr[\varrho ^{1/2},K]^{2}.
\label{skew}
\end{equation}%
It is especially noted that \textit{K}-coherence depends not only on the
measured state $\varrho $ but also on the observable $K$. If $K$ is
degenerate, $I\left( \varrho ,K\right) $ only detects the coherence in the
non-degenerate subspace of $K$.

Now let's turn to the SD $\Delta A$ of an observable $A$ subject to a pure
state $\varrho =\left\vert \psi \right\rangle \left\langle \psi \right\vert $%
. It is easy to show that 
\begin{equation}
\Delta _{\psi }A=\sqrt{I\left( \varrho ,A\right) }=\sqrt{-\frac{1}{2}%
Tr[\varrho ^{1/2},A]^{2}}  \label{eql}
\end{equation}%
with $\varrho =\left\vert \psi \right\rangle \left\langle \psi \right\vert $
and $\left[ \cdot ,\cdot \right] $ denoting the commutation relation. This
relation directly shows that the equivalence between the SD and the \textit{K%
}-coherence for any an observable $A$ on a pure state. Thus one will easily
obtain how the \textit{K}-coherence is distributed among the superposed
components.

\noindent\textbf{Corollary 1.-} \textit{For the superposition state} $%
\left\vert \psi \right\rangle $ \textit{defined in Theorem 1, the \textit{K}%
-coherence subject to the observable} $A$ \textit{is bounded by}%
\begin{equation}
\mathcal{B}_{L}\leq \left\Vert \left\vert \psi \right\rangle \right\Vert
^{2}I\left( \left\vert \tilde{\psi}\right\rangle \left\langle \tilde{\psi}%
\right\vert ,A\right) \leq \mathcal{B}_{U}  \label{bc}
\end{equation}%
\textit{where} $\mathcal{B}_{L},\mathcal{B}_{U}$ \textit{have the same form
in Theorem 1 but all the} $\Delta _{\psi _{i}}A$ \textit{should be replaced
by their corresponding} $I\left( \left\vert \psi _{i}\right\rangle
\left\langle \psi _{i}\right\vert ,A\right) $.

\noindent\textbf{Proof.} This is a direct result of Eq. (\ref{eql}).\hfill $%
\blacksquare $

\section{Bounds on the incompatibility}

As mentioned at the beginning, the SD is the important ingredient for the
remarkable Heisenberg uncertainty principle (HUP) \cite{Busch}. However, the
HUP is expressed in terms of the product of the SDs of two observables, so
it could lead to a trivial bound even though two incompatible observables
are taken into account \cite{wat}. Recently, Maccone and Pati \cite{pati}
have raised another type of uncertainty relation by considering the sum of
the SDs of two observables. They showed that the uncertainty relations would
not get a trivial bound at any rate. It is natural that the nontrivial bound
usually shows whether the considered observables are compatible or not. In
fact, we would like to emphasize that the exact value of the sum of the SDs
(or the exact value of the uncertainty) just signals to what degree the
considered observables are incompatible. In this sense, we can define the
incompatibility \cite{expl} of two operators $A$ and $B$ subject to a given
normalized state $\left\vert \psi\right\rangle $ with $\left\vert \tilde{\psi%
}\right\rangle= \frac{\left\vert \psi\right\rangle}{\left\Vert\psi\right\Vert%
}$ as 
\begin{equation}
U_{\psi }\left( A,B\right) =\Delta _{\tilde{\psi}}^{2}A+\Delta _{\tilde{\psi}%
}^{2}B.  \label{inc}
\end{equation}%
It is obvious that $U_{\psi}\left( A,B\right) =0$ means that $A$ and $B$ can
be simultaneously approximately measured on the state $\left\vert \tilde{\psi%
}\right\rangle $. The larger $U_{\psi }\left( A,B\right) $ is, the more
incompatible $A$ and $B$. In addition, a reasonable (lower) bound for $%
U_{\psi }\left( A,B\right) $ could form an uncertainty relation. Now let's
consider when $\left\vert \psi \right\rangle $ is a superposition state, how
the incompatibility can be distributed among every superposed component.

\noindent \textbf{Corollary 2.-}\textit{Let }$\left\vert \psi \right\rangle $%
\textit{\ be defined in Theorem 1}, \textit{the incompatibility of two
observable }$A$ \textit{and} $B$ $U_{\psi }\left( A,B\right) $ \textit{is
bounded as} 
\begin{equation}
\tilde{\mathcal{B}}_{L}\leq \left\Vert \left\vert \psi \right\rangle
\right\Vert ^{2}U_{\tilde{\psi}}\left( A,B\right) \leq \tilde{\mathcal{B}}%
_{U}  \label{ib1}
\end{equation}%
\textit{with } 
\begin{equation}
\tilde{\mathcal{B}}_{L}=\max \{\tilde{b}_{L},0\},  \label{lbe}
\end{equation}%
\textit{where } 
\begin{equation}
\tilde{b}_{L}=\sum\limits_{i=1}^{N}\left\vert \alpha _{i}\right\vert
^{2}U_{\psi _{i}}\left( A,B\right) -\sum\limits_{X=A,B}E_{+}(X)-\tilde{F}%
\left( A,B\right)  \label{lbcc}
\end{equation}%
\textit{and} 
\begin{equation}
\tilde{\mathcal{B}}_{U}=\sum\limits_{i=1}^{N}\left\vert \alpha
_{i}\right\vert ^{2}U_{\psi _{i}}\left( A,B\right)
-\sum\limits_{X=A,B}E_{-}(X)+\tilde{F}\left( A,B\right)  \label{ubb}
\end{equation}%
\textit{with} 
\begin{eqnarray}
\tilde{F}\left( A,B\right) &=&\sum\limits_{i,j=1,i<j}^{N}2\left\vert \alpha
_{i}\alpha _{j}\right\vert \sqrt{U_{\psi _{i}}\left( A,B\right)
+\left\langle A\right\rangle _{i}^{2}+\left\langle B\right\rangle _{i}^{2}} 
\notag \\
&&\times \sqrt{U_{\psi _{j}}\left( A,B\right) +\left\langle A\right\rangle
_{j}^{2}+\left\langle B\right\rangle _{j}^{2}}  \label{fab}
\end{eqnarray}%
\textit{and }$E_{\pm }(A)$ \textit{defined as Theorem 1.}

\textbf{Proof.} For the observable $A$ and the state $\left\vert \psi
\right\rangle $, we can obtain, from Theorem 1, the bounds of SD of $A$ as%
\begin{equation}
\mathcal{B}_{L}(A)\leq \left\Vert \left\vert \psi \right\rangle \right\Vert
^{2}\Delta _{\tilde{\psi}}^{2}A\leq \mathcal{B}_{U}(A)  \label{pas}
\end{equation}%
where $\mathcal{B}_{L}(A)$ and $\mathcal{B}_{U}(A)$ with the same form as $%
\mathcal{B}_{L}$ and $\mathcal{B}_{U}$ just show that the bounds corresponds
to the observable $A$. Similary, for $B$ we have 
\begin{equation}
\mathcal{B}_{L}(B)\leq \left\Vert \left\vert \psi \right\rangle \right\Vert
^{2}\Delta _{\tilde{\psi}}^{2}B\leq \mathcal{B}_{U}(B)  \label{pbs}
\end{equation}%
Sum Eq. (\ref{pas}) and Eq. (\ref{pbs}), one will arrive at 
\begin{eqnarray}
\mathcal{B}_{L}(A)+\mathcal{B}_{L}(B) &\leq &\left\Vert \left\vert \psi
\right\rangle \right\Vert ^{2}U_{\tilde{\psi}}\left( A,B\right)  \notag \\
&\leq &\mathcal{B}_{U}(A)+\mathcal{B}_{U}(B)  \label{pab}
\end{eqnarray}%
Substitute Eq. (\ref{lbc}) and Eq. (\ref{ub}) into Eq. (\ref{pab}), it can
be found that%
\begin{eqnarray}
&&F(A)+F(B)  \notag \\
&=&\sum\limits_{i,j=1,i<j}^{N}2\left\vert \alpha _{i}\alpha _{j}\right\vert 
\sqrt{\left( \Delta _{_{\psi _{i}}}^{2}\left( A\right) +\left\langle
A\right\rangle _{i}^{2}\right) \left( \Delta _{_{\psi _{j}}}^{2}\left(
A\right) +\left\langle A\right\rangle _{j}^{2}\right) }  \notag \\
&&+\sum\limits_{i,j=1,i<j}^{N}2\left\vert \alpha _{i}\alpha _{j}\right\vert 
\sqrt{\left( \Delta _{_{\psi _{i}}}^{2}\left( B\right) +\left\langle
B\right\rangle _{i}^{2}\right) \left( \Delta _{_{\psi _{j}}}^{2}\left(
B\right) +\left\langle B\right\rangle _{j}^{2}\right) }  \notag \\
&\leq &\sum\limits_{i,j=1,i<j}^{N}2\left\vert \alpha _{i}\alpha
_{j}\right\vert \sqrt{U_{\psi _{i}}\left( A,B\right) +\left\langle
A\right\rangle _{i}^{2}+\left\langle B\right\rangle _{i}^{2}}  \notag \\
&&\times \sqrt{U_{\psi _{j}}\left( A,B\right) +\left\langle A\right\rangle
_{j}^{2}+\left\langle B\right\rangle _{j}^{2}}  \notag \\
&=&\tilde{F}\left( A,B\right)  \label{pfinal}
\end{eqnarray}%
which is based on the Cauchy-Schwarz inequality. Thus one can easily find
that the upper and lower bounds are given just as Eq. (\ref{lbcc}) and Eq. (%
\ref{ubb}). The proof is completed.\hfill $\blacksquare $

\section{Conclusion and discussion}

We have derived an upper bound and a lower bound, respectively, for the SD
of a superposition state in terms of the SDs of the superposed components.
This lets us well understand how the SD is distributed among every
superposed component. Numerical examples are given to test the tightness of
the bounds. It is shown that such bounds can be well suitable for the
distribution of the coherence of superposition states, since the coherence
and the SD have the consistent form of definition based on the skew
information. As a further connection with Heisenberg uncertainty principle,
we suggest an alternative definition of incompatibility of two observables
subject to a given state. Considering the superposition of state, we also
study how we can evaluate the incompatibility of two observables subject to
a superposition state in terms of the incompatibilities of every superposed
component.

Since the SD is not the unique quantification of the uncertainty of the
repeated measurement outcomes, it easily comes to our mind that the various
entropy-based measure are also good candidates \cite{winter, Deutsch,
Maassen,miyao,Coles,Rud,ZJ}. In particular, they don't include the
contribution of the eigenvalues of the observable. So how these types of
measures are distributed among the different superposed components and what
novel results could be implied in these relations deserve us forthcoming
efforts.

\section{Acknowledgements}

This work was supported by the National Natural Science Foundation of China,
under Grant No.11375036 and 11175033, the Xinghai Scholar Cultivation Plan
and the Fundamental Research Funds for the Central Universities under Grant
No. DUT15LK35 and No. DUT15TD47.

\end{document}